%% file: leop_optical_astroph.tex
\documentclass[iop]{emulateapj}

\usepackage{longtable}

\newcommand{\arcm}{\ifmmode {' }\else $' $\fi}
\newcommand{\arcs}{\ifmmode {'' }\else $'' $\fi}
\newcommand{\etal}{{\it et\thinspace al.}\ }
\newcommand{\kms}{km\thinspace s$^{-1}$}
\newcommand{\simlt}{\ {\raise-.5ex\hbox{$\buildrel<\over\sim$}}\ }
\newcommand{\hi}{\ion{H}{1}}
\newcommand{\hii}{\ion{H}{2}}

\def\be{\begin{equation}}
\def\ee{\end{equation}}

\slugcomment{Accepted to AJ on 2 April 2013}

\shortauthors{Rhode et al.} \shorttitle{New Dwarf Galaxy: Optical Imaging}

\begin{document}

\title{ALFALFA Discovery of the Nearby Gas-Rich Dwarf Galaxy Leo~P. \\ II. Optical Imaging Observations}

\author{Katherine L. Rhode\altaffilmark{1}, John J. Salzer\altaffilmark{1,2}, Nathalie C. Haurberg\altaffilmark{1}, Angela Van Sistine\altaffilmark{1,2}, Michael D. Young\altaffilmark{1}, Martha P. Haynes\altaffilmark{3}, Riccardo Giovanelli\altaffilmark{3}, John M. Cannon\altaffilmark{4}, Evan D. Skillman\altaffilmark{5}, Kristen B. W. McQuinn\altaffilmark{5}, \& Elizabeth A. K. Adams\altaffilmark{3} } 

\altaffiltext{1}{Department of Astronomy, Indiana University, 727 East Third
  Street, Bloomington, IN 47405.  {\it e--mail:}  rhode@astro.indiana.edu, slaz@astro.indiana.edu}
  
\altaffiltext{2}{Visiting Astronomer, Kitt Peak National Observatory,
National Optical Astronomy Observatory, which is operated by the
Association of Universities for Research in Astronomy (AURA)
under cooperative agreement with the National Science Foundation.}

\altaffiltext{3}{Center for Radiophysics and Space Research, Space Sciences Building,
Cornell University, Ithaca, NY 14853. {\it e--mail:} riccardo@astro.cornell.edu,
haynes@astro.cornell.edu, betsey@astro.cornell.edu}

\altaffiltext{4}{Department of Physics and Astronomy,
Macalester College, Saint Paul, MN 55105. {\it e--mail:} jcannon@macalester.edu}

\altaffiltext{5}{Minnesota Institute for Astrophysics,
University of Minnesota, Minneapolis, MN 55455. {\it e--mail:} skillman@astro.umn.edu, kmcquinn@astro.umn.edu}

\begin{abstract}
We present results from ground-based optical imaging of a low-mass
dwarf galaxy discovered by the ALFALFA 21-cm \hi\ survey.  Broadband
(BVR) data obtained with the WIYN 3.5-m telescope at Kitt Peak
National Observatory (KPNO) are used to construct color-magnitude
diagrams of the galaxy's stellar population down to V$_o$ $\sim$ 25.  We
also use narrowband H$\alpha$ imaging from the KPNO 2.1-m telescope to
identify an \hii\ region in the galaxy.  We use these data to constrain the 
distance to the galaxy to be between 1.5 and 2.0 Mpc.  This 
places Leo~P within the Local Volume but beyond the Local Group.  Its 
properties are extreme: it is the lowest-mass system known that contains 
significant amounts of gas and is currently forming stars.   
\end{abstract}


\section{Introduction}
\label{section:introduction}

The number of known galaxies in and around the Local Group (LG) is
increasing with time, as new dwarf galaxies are discovered
serendipitously in wide-field surveys and via dedicated searches
(e.g., Armandroff, Davies, \& Jacoby 1998, Karachentsev et al.\ 2000,
Willman et al.\ 2005, Whiting et al.\ 2007).  The 1998 review article
by Mateo on dwarf galaxies in the LG lists $\sim$38 dwarfs in its
census.  A recent review by McConnachie (2012) summarized the observed
properties of nearby dwarf galaxies and listed $\sim$70 ``definite''
or ``very likely'' LG dwarfs and a total of more than 90 dwarf
galaxies within 3~Mpc of the Sun.
McConnachie notes that not only has the number of nearby dwarf
galaxies essentially doubled over roughly the last decade, but
improvements in observing capabilities over the same time period have
enabled us to study the star formation histories and other properties
of dwarf galaxies in much more detail (e.g., Tolstoy et al.\ 2009 and
references therein).  The newly-identified low-mass systems give us a
more complete census of the 
Local Volume and
provide a testing ground for ideas and theories that are fundamental
to many areas of astrophysics --- ideas about chemical evolution, star
formation, stellar feedback processes, galaxy evolution, hierarchical
galaxy assembly, and dark matter (e.g., Tolstoy et al.\ 2009).

The majority of recently discovered dwarf galaxies located within the
LG or in the immediate vicinity are gas-poor dwarf spheroidal (dSph)
galaxies and/or ultra-low-luminosity systems that have been detected
as small surface brightness enhancements or slight overdensities of
stars (e.g., Martin et al.\ 2006, Belokurov et al.\ 2007, Bell,
Slater, \& Martin 2011).  Only a few have detectable amounts of gas
(e.g., Phoenix: Canterna \& Flower 1977, Young et al.\ 2007; Leo T:
Irwin et al.\ 2007), and these tend to be located relatively far from
the two giant members of the LG, Andromeda and the Milky Way.  In
contrast to the small number of dwarfs with gas, wide area 21-cm radio
surveys have cataloged hundreds of \hi\ clouds with velocities that
potentially place them in and around the LG
(Blitz et al.\ 1999; Braun \& Burton 1999; Giovanelli et al.\ 2010).  In the 
current paper, we present optical observations of one such object, the
gas-rich galaxy AGC 208583 ( = Leo~P), discovered by the ALFALFA \hi\ survey (Giovanelli et
al.\ 2005; Haynes et al.\ 2011) and described in Giovanelli et al.\ (2013).

We have obtained ground-based narrowband H$\alpha$ and broadband BVR
images of Leo~P and made photometric measurements of the resolved
stellar population detected in the data. We use these optical data to
construct color-magnitude diagrams (CMDs) of the galaxy's stellar component.
These CMDs are used to constrain the distance to the galaxy and show that Leo~P 
is likely to be located between 1.5 and 2.0 Mpc from the Milky Way.  In addition, 
we highlight  the unusual nature of this system by quantifying its physical properties
(e.g., size, H$\alpha$ luminosity, $V$-band luminosity, colors,
and \hi\ \and stellar mass).

\begin{figure*}
\centering
\includegraphics[width=5.8in]{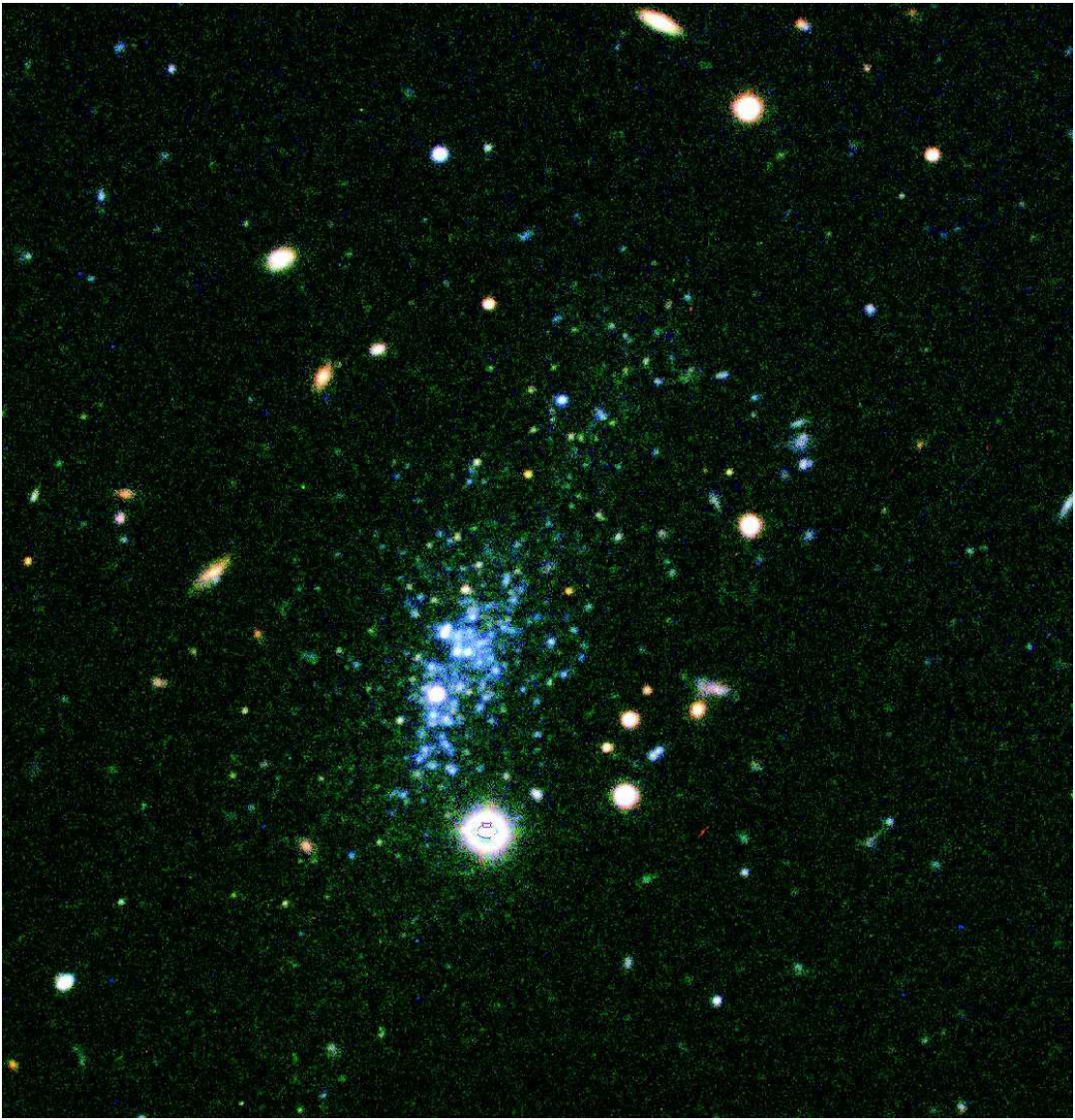}
\caption{BVR color composite image of Leo~P obtained with the WIYN
  3.5-m telescope.  The FOV of this image is 2.4\arcmin\ by
  2.5\arcmin\ and the orientation is N-up, E-left.  The lower
  (southern) portion of Leo~P is dominated by a clump of blue
  stars, indicating very recent star formation has occurred.
  The brightest object in Leo~P (located above and to the left of
  the bright foreground star) is an HII region that appears to be
  photoionized by an O or B-type star or stellar association.  The upper portion of the galaxy is
  very low surface brightness but includes a number of redder stars,
  presumably RGB members in Leo~P.  The total size of the galaxy at
  this sensitivity level is $\sim$90\arcsec .}
\label{bvr color image}
\end{figure*}

\section{Observations}
\label{section:observations}

\subsection{Broadband BVR Imaging}

The WIYN Observatory\footnote{The WIYN Observatory is a joint facility
  of the University of Wisconsin-Madison, Indiana University, Yale
  University, and the National Optical Astronomy Observatory.} 3.5-m
telescope was used on 26 March 2012 to image the field located
at the position of an ALFALFA detection at 10h 21m 45.0s +18d 05m 01s.
These observations revealed an optical counterpart with a resolved
stellar component.
All observations were obtained with the Minimosaic camera through
standard BVR broadband filters under photometric conditions.  Total
integration times of 30 minutes in B, 24 minutes in V, and 20 minutes
in R were achieved, split between two images for each filter.  Stellar
point-spread function (PSF) measurements ranged between
0.6\arcsec\ and 0.8\arcsec\ for the sequence of observations.  Images
of Landolt (1992) standard stars taken before and after the
observations of Leo~P provided photometric calibrations.  Errors on
the photometric coefficients were $<$0.01~mag.  Data reduction
followed normal practices.

A composite three-color image of  Leo~P is shown in 
Figure 1.  The image is oriented N-up E-left.  The galaxy is resolved into 
stars, with a strong concentration of the brightest and bluest stars
appearing in the southern portion.  As we discuss in \S 3, these may be
upper main-sequence stars with  B and A spectral types.  A number of
fainter and redder stars are also present that are presumably red
giants from an older population.  These redder stars appear to be
uniformly distributed throughout the galaxy, although in the northern,
lower-surface-brightness region they are the only stars present.  

The angular diameter of the optical portion of the galaxy is $\sim$90
arcsec. The size of the galaxy could not easily be constrained via
surface brightness measurements, so the diameter is based on the
apparent extent of the stellar point sources. To measure the extent,
we used the list of all point sources in the image with color errors
less than 0.25 mag (see Section 3).  We assigned
each of the point sources to one of a series of 0.05-arcmin-wide concentric
annuli centered on the approximate midpoint of the galaxy
light distribution. An effective area was calculated for each annulus
that excluded portions of the annulus that extended off the edge of the
image.  The number of point sources in each annulus was then divided
by the effective area of the annulus to produce a radial profile
(surface density of stars versus projected radius) of the stellar
distribution.  The surface density falls to the level of the
background stellar density at $~$0.725 arcmin from the galaxy center; from
this we estimate that the apparent diameter of the
galaxy is 87 arcsec with an uncertainty of $\pm$3 arcsec (the width of
one annulus).  The photometric properties of the stars in Leo~P are
discussed in detail in the next section.

\subsection{H$\alpha$ Imaging}

Narrowband H$\alpha$ images of Leo~P were obtained with the KPNO 2.1-m
telescope on 20 March 2012.  The observations consist of two 15-minute
exposures taken with a narrowband filter (central $\lambda$ =
6573~\AA, bandwidth = 67~\AA) sandwiched around a single 3-minute
R-band image that was used for continuum subtraction.  An additional
15-minute R-band image was taken to provide a deeper exposure of the
system.  Standard image processing steps were utilized to produce a
continuum-subtracted image of Leo~P in the light of the H$\alpha$
emission line (Figure 2).  Observations of spectrophotometric standard
stars (Oke \& Gunn 1983) allowed us to calibrate the observed
narrowband flux.

Figure 2 shows that Leo~P possesses a single \hii\ region located
near the southern end of the galaxy.  The \hii\ region is associated
with a bright, blue star and appears as the single brightest object in
the system.   The \hii\ region is spatially resolved, and has an apparent 
diameter of 1.20 $\pm$ 0.05 arcsec.  The observed emission-line flux 
is (1.71 $\pm$ 0.03) 
$\times$ 10$^{-14}$ erg/s/cm$^2$.   Our narrow-band data are quite 
sensitive, with a 5$\sigma$ point source detection limit of
$\sim$5.0 $\times$ 10$^{-17}$ erg/s/cm$^2$.

\begin{figure}
\centering
\includegraphics[width=3.4in]{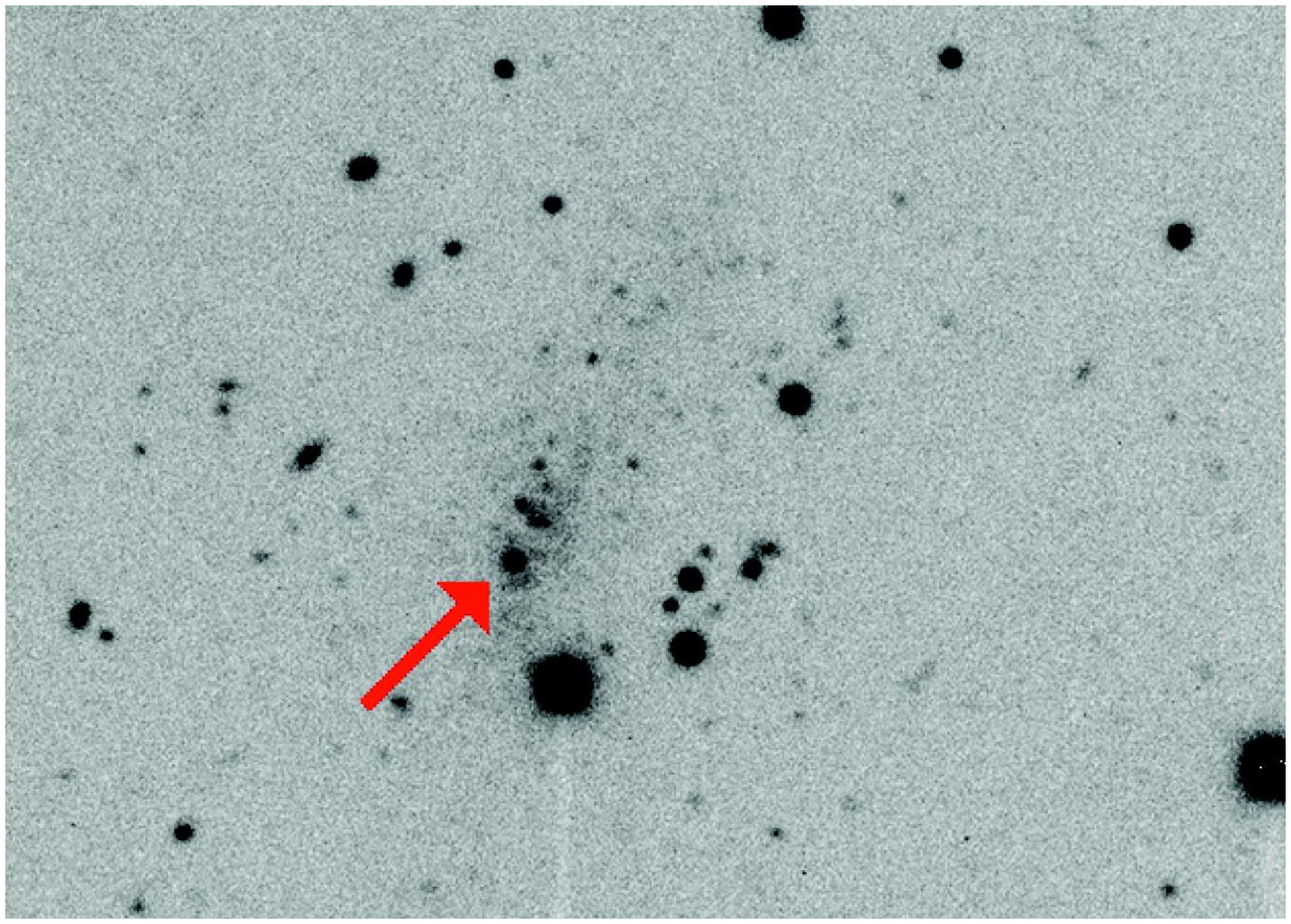}\hskip 0.1in \includegraphics[width=3.4in]{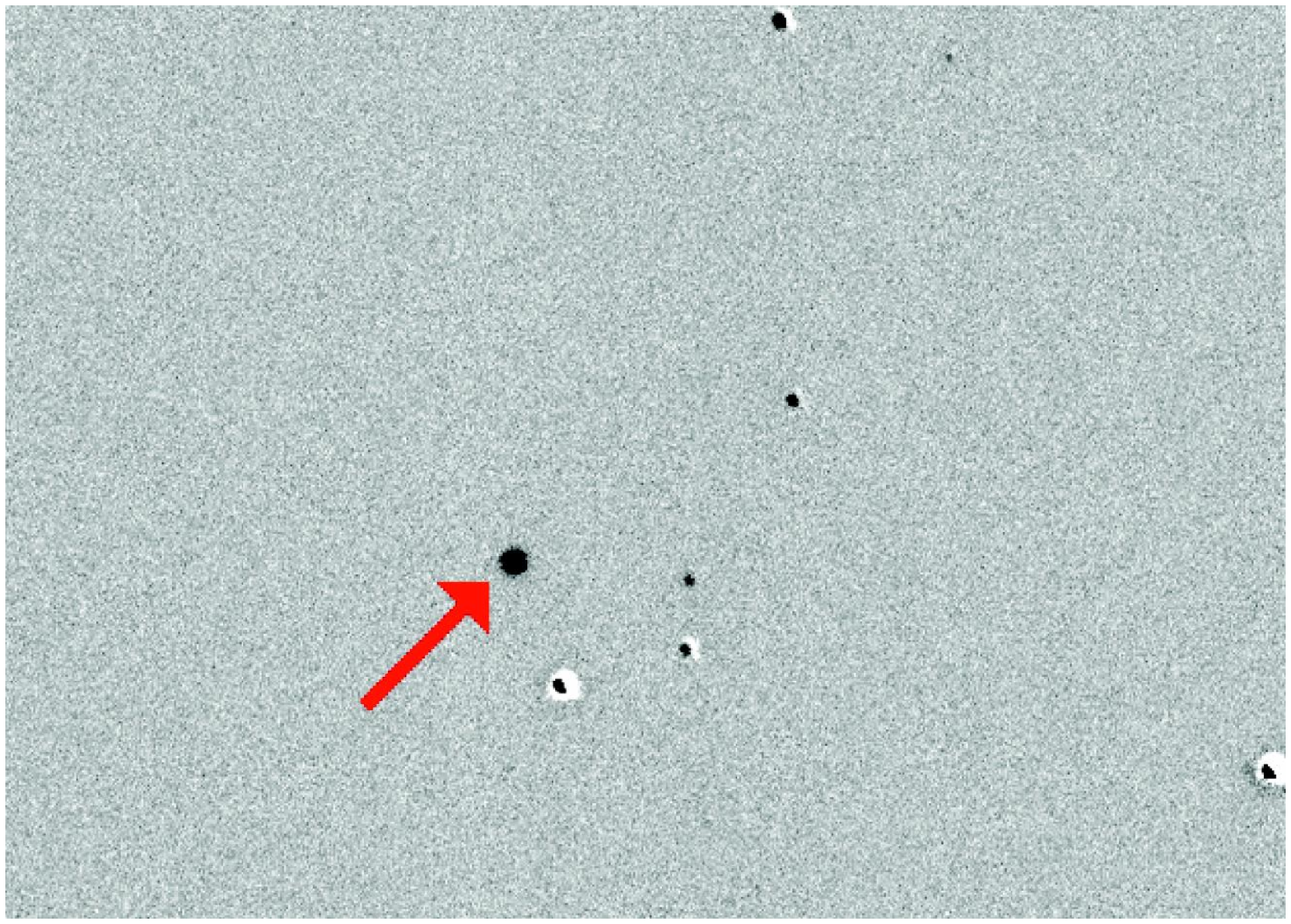}
\caption{Narrowband imaging of Leo~P.  The left panel is an R-band
  continuum image and the right panel is a continuum-subtracted
  H$\alpha$ image, both taken with the KPNO 2.1m telescope.  The
  images are oriented N-up, E-left.  There is a single HII region
  present in Leo~P (location marked with an arrow in both panels), 
  with an integrated flux of 1.7 $\times$ 10$^{-14}$
  erg/s/cm$^2$.  The points of emission to the right of the HII region
  in the right-hand panel are caused by imperfect continuum
  subtraction in some of the brighter stars.}
\label{halpha image}
\end{figure}

\section{Photometry and CMDs of Leo~P}

We utilized the PSF-fitting photometry software DAOPHOT (Stetson 1987,
1990) to measure the brightnesses of the stars in the broadband BVR images 
obtained with the WIYN 3.5-m telescope.
A model PSF was constructed using a selection of foreground stars as
well as relatively bright, isolated stars believed to be members of
Leo~P.  Photometry was performed on each broadband image using a list
of stars selected from a combined R and V image and allowing the
program to redetermine the precise position of the centroids in each
frame independently. We calculated and applied aperture corrections to
the results and calibrated the apparent magnitudes using the
photometric solution derived from the Landolt standard star
observations.  As a check of the PSF-fitting photometry, we measured
the fluxes from isolated single stars using aperture photometry
methods.  These checks verified the integrity of the DAOPHOT results.

Color-magnitude diagrams (CMDs) of the resolved stars in Leo~P are
shown in Figure 3.   We present CMDs in two colors: (B$-$V)$_o$ (Figure 3a) and
(V$-$R)$_o$ (Figure 3b).   All stars brighter than V$_o$ $\sim$ 25.0 have been measured,  
but only stars with color errors less than 0.25 mag are included in the plots 
(N=82 for the (B$-$V)$_o$ plot and N=77 for (V$-$R)$_o$).  
The positions and photometric measurements of the 82 stars included in Figure 3a 
are listed in Table 1.   All photometric quantities reported in this paper are 
corrected for Galactic absorption using E(B$-$V) = 0.026 from Schlegel et al.\ (1998).

Unfortunately, our B photometry is relatively shallow, so the photometric
uncertainties for the redder stars in Figure 3a are large.   Likewise the R data are
not as deep as the V data, leading to larger errors for the bluer stars in Figure 3b.
Because of the differing depths of the images taken through the B, V and R filters,
the main sequence (MS) in Leo~P is better defined in the B$-$V plot, while the red 
giant branch has better definition in the V$-$R CMD.
We performed a series of artificial star tests in order to quantify the point-source 
detection limits of the WIYN images. Thirty artificial stars with magnitudes within 
0.2 mag of a fiducial value and the appropriate point spread function (PSF) were 
added to each of the combined images (B, V, and R).  We then ran detection 
software to determine what fraction of these artificial objects would be recovered.  
We repeated this step in 0.2-magnitude intervals over a range of 4--5 magnitudes for 
each image.  The result was a series of curves that quantify the completeness as 
a function of magnitude in each of the combined images.  Dashed lines in Figure 3 
(and all subsequent CMD plots) show the 50\% completeness levels of our data.

\begin{longtable*}{rccccrcrc}
\tabletypesize{\tiny}
\tablecaption{WIYN Broadband Photometry of Resolved Stars in Leo~P \label{tab:phot}}
\tablehead{
\colhead{\#}& \colhead{RA (2000)}& \colhead{Dec (2000)}& \colhead{$V_o$}& \colhead{$\sigma_{V}$}& \colhead{$(B-V)_o$}& \colhead{$\sigma_{B-V}$} & \colhead{$(V-R)_o$}& \colhead{$\sigma_{V-R}$}
}
\startdata
\input{table_1.dat}
\enddata
\end{longtable*}

Two features are immediately apparent in the CMDs.  First, a sequence of 
bright, blue stars is present in Leo~P.  
The six brightest stars in our CMDs (V$_o$ brighter than $\sim$22.5) all have
(B$-$V)$_o$ between $-$0.10 and $-$0.21.  The brightest object is the star
located at the center of the single \hii\ region (the first star in
Table 1; hereafter Star 1).  The photometry of this star has been
corrected for nebular contamination using the emission-line spectrum
obtained by our group and described below. 
Second, a red giant branch (RGB) appears to be
present at (B$-$V)$_o$ $\sim$ 0.7 -- 1.3 and (V$-$R)$_o$ $\sim$ 0.4 -- 0.8,
beginning at V$_o$ $\sim$ 23.3 and extending to fainter magnitudes.
Unfortunately, the apparent location of the upper end of the RGB coincides with
the onset of larger photometric errors in our data, making a secure identification
of the RGB tip problematic.
We return to this issue in \S 4.1.1 below.

\begin{figure}
\centering
\includegraphics[width=3.4in]{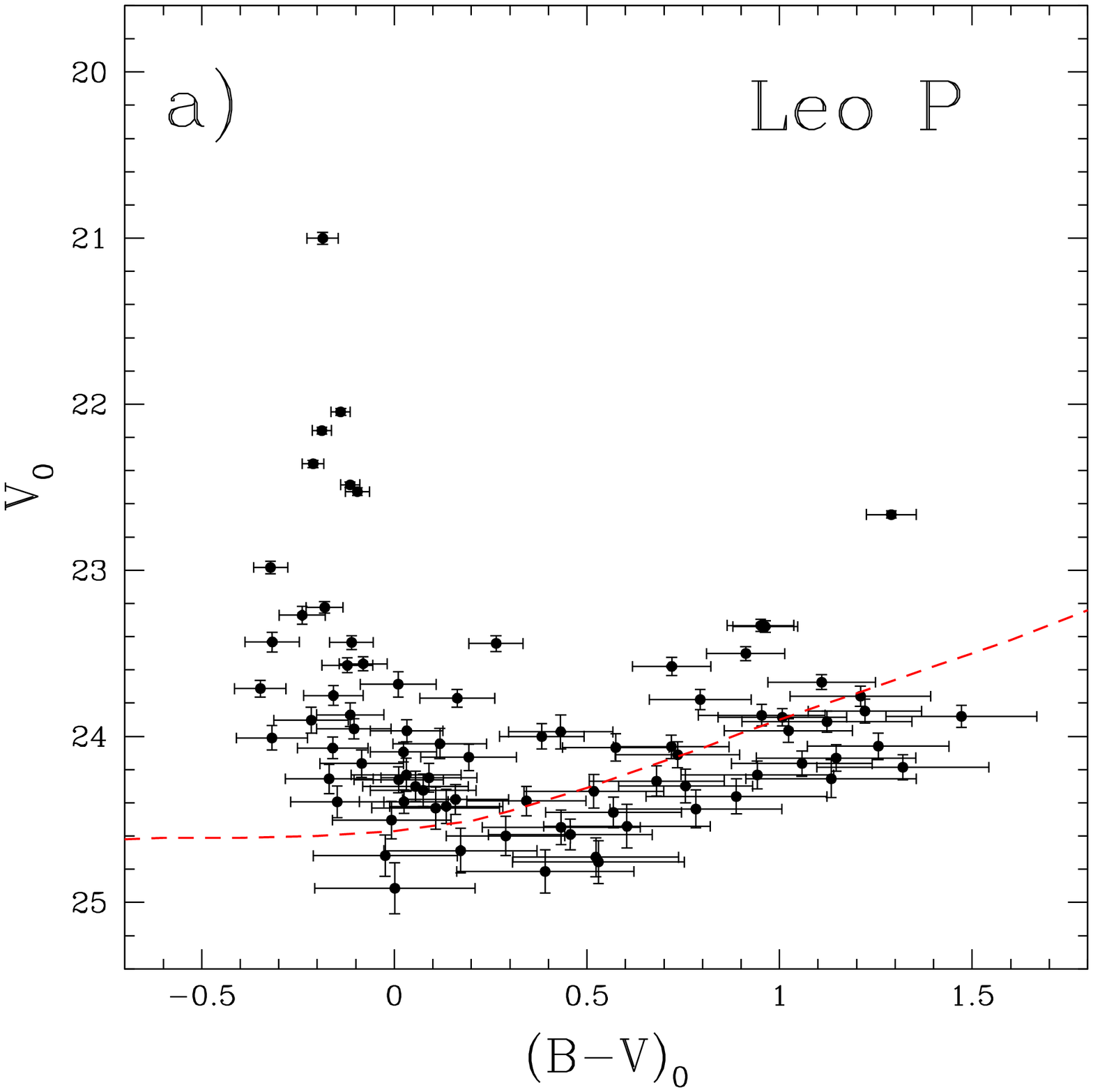}\hskip 0.1in \includegraphics[width=3.4in]{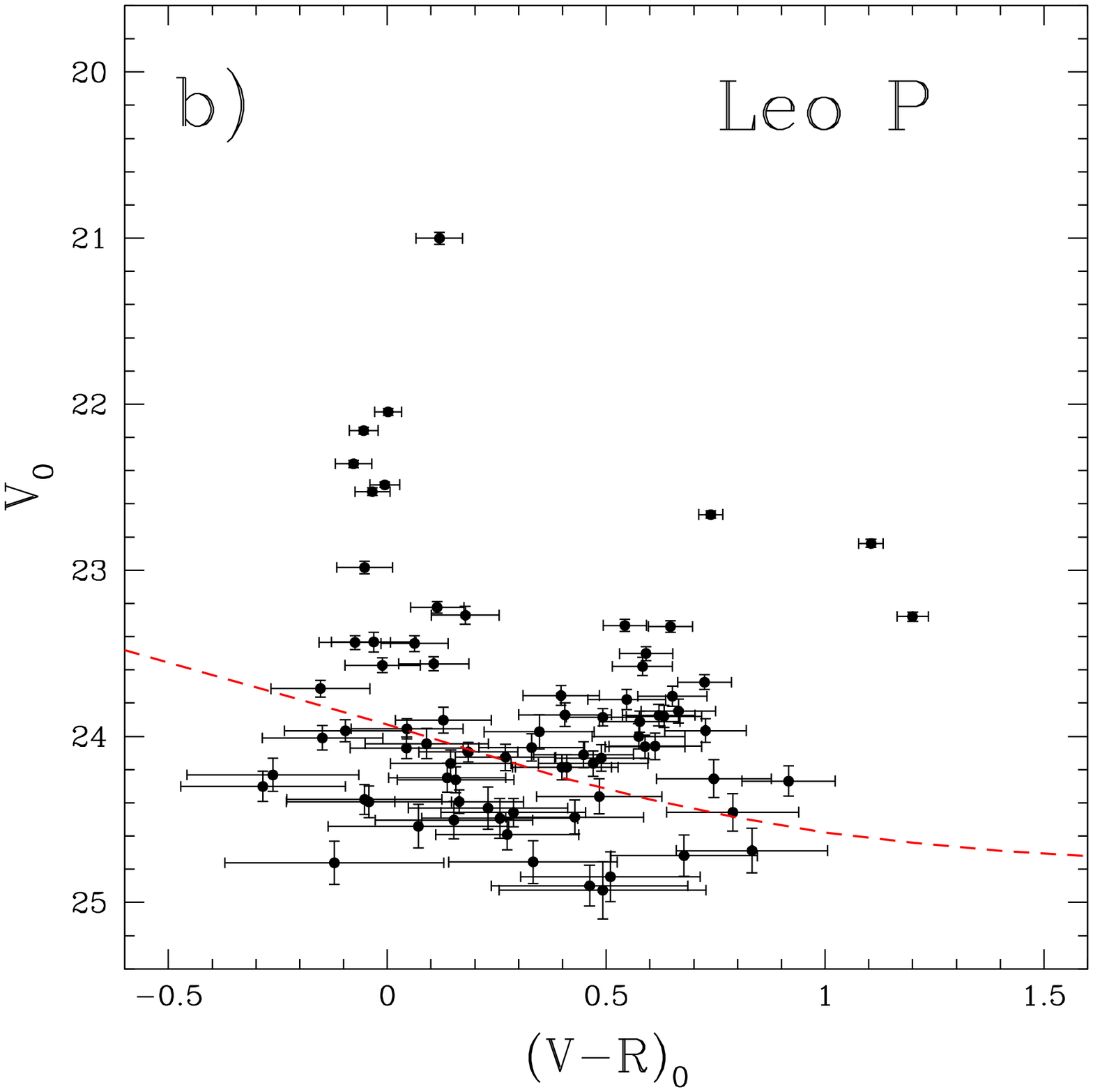}
\caption{Color-magnitude diagrams of the resolved stars in Leo~P.  The
  left plot shows (B$-$V)$_o$ color, while the right shows (V$-$R)$_o$.  Due to
  the varying depths of the BVR images, the B$-$V CMD shows better
  definition of the MS, whereas the V$-$R plot shows the RGB more
  clearly.  The dashed curves indicate the 50\% completeness level for
  our data.}
\end{figure}

Figure 4 presents the results of an analysis of the PSF photometry for
the entire field of view included in our images of Leo~P.  The upper
left plot (panel a) shows the data for the stars in Leo~P; this is
identical to Figure 3a.  In the upper right plot (panel b) we show the
corresponding CMD for the entire CCD frame, with the area covered by
Leo~P excluded.  This plot shows the locations of the foreground stars
and unresolved galaxies in the Leo~P field (all obviously extended
galaxies have been removed).  When the two panels are compared, it is
immediately obvious that the bright, blue stars in the Leo~P
area are completely absent from  the full-field CMD.  There are 13 stars
in the Leo~P area with (B$-$V)$_o$ colors less than 0.0 and with V$_o$
magnitudes brighter than 23.6, whereas there is not a single star in the
rest of the image that falls within this range, despite the fact that
the area covered in panel (b) is 51.3 times larger.  Hence, there can be
no question that the bright blue stars seen in Figure 1 are all
associated with Leo~P.

The lower two panels of Figure 4 show the CMDs for two random fields
located in our CCD frames.  The areas of the two regions are identical
to that covered by Leo~P (0.875 sq. arcmin).  The fields are centered
on the same declination as the Leo~P area, but are offset by $\sim$3
arcmin in RA, one to the east and one to the west.  These fields can
be thought of as representing the foreground (or background)
contamination present in the Leo~P CMD.  The two fields contain 6 and
9 stars, respectively, but only one star per field with V$_o$ magnitudes
brighter than 23.8.   This suggests that foreground contamination of the 
Leo~P CMD, especially at V$_o$ brighter than $\sim$24.0, is minimal. 
We quantify the contamination level of the Leo~P CMD in the 
RGB region of the diagram in \S 4.1.1.

\begin{figure}
\centering
\includegraphics[width=3.4in]{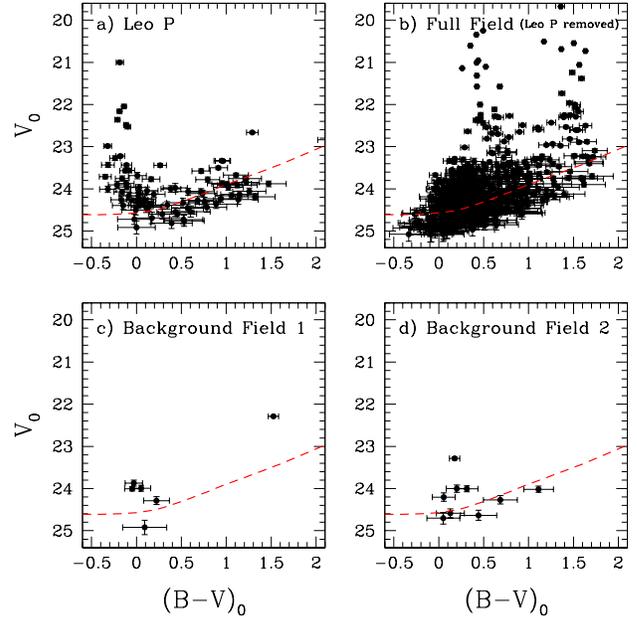}
\caption{CMDs for (a) Leo~P, with the same data as shown in Figure 3a; (b) the full CCD image that contains Leo~P, but with the stars from the Leo~P region excluded; (c) and (d) two background comparison fields with the same area as the Leo~P photometry but located far from the galaxy.   The strong upper MS identified in Leo~P is totally absent from the surrounding field, as indicated by the complete lack of stars with (B$-$V)$_o$ $\le$ 0.0 and brighter than V$_o$ $\sim$ 23.6 in panel (b).  The two lower panels indicate that foreground contamination of the Leo~P CMD is likely to be minimal.}
\label{full field CMD}
\end{figure}

In Figure 5 we replot the data shown in Figure 3a with model
isochrones overlaid.  The isochrones are from the Padova group
(Girardi et al.\ 2002) for a 10 Myr old population with a metallicity
(Z) of 0.0004 (1/50th solar).   The isochrones are plotted for four 
different distances: 0.5 Mpc (green), 1.0 Mpc (red),
1.5 Mpc (blue), and 2.0 Mpc (magenta).  The values for age and Z 
were selected to be reasonable matches for Leo~P (see \S 4.2), but 
in fact the locations of the isochrones change only slightly if different 
(but similar) ages or metallicities are used.   To illustrate this last point,
we plot a single 20 Myr isochrone for D = 1.5 Mpc (dashed blue line).

\begin{figure}
\centering
\includegraphics[width=3.4in]{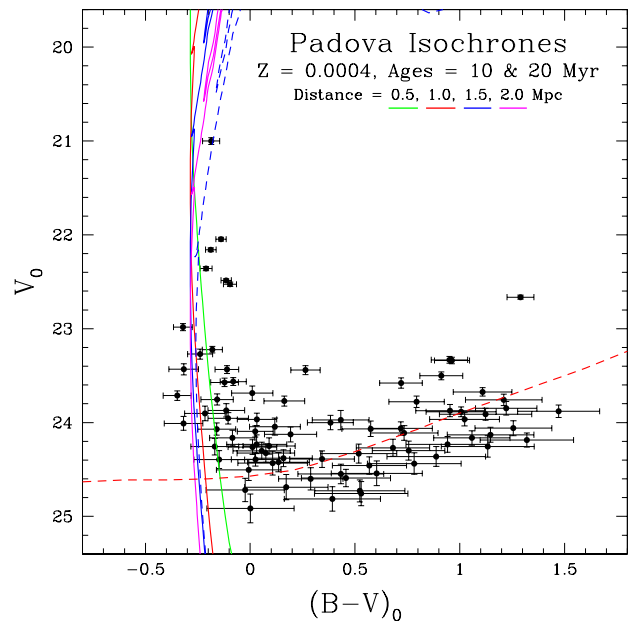}
\caption{CMD constructed from PSF photometry of the stars in Leo~P.   The data points and 50\% completeness line are the same as shown in Figure 3a.   Overlaid on the MS stars are isochrones from the Padova group (Girardi et al.\ 2002) for a 10 Myr old population with Z of 2\% solar and for a range of assumed distances as indicated in the legend (0.5 - 2.0 Mpc).   The blue dashed isochrone (for a distance of 1.5 Mpc) shows the effect of increasing the age of the young stellar population to 20 Myr.   The blue and magenta isochrones bracket our preferred distance estimate based on the apparent location of the tip of the RGB.   \protect\label{fig:psf bv cmd}}
\end{figure}

We stress that these young-population isochrones cannot be used effectively 
to constrain the distance to Leo~P, since the only portion of the CMD that
is well defined is the upper MS, where the isochrones are nearly vertical.  
The only distinction between the Girardi et al.\ (2002) isochrones of different 
distances in the upper-MS region is a small color shift.  By the time the MS starts 
to flatten slightly at colors redder than (B$-$V)$_o$ = 0.0, the photometric errors 
for the MS stars are far too large to allow us to discriminate between the various 
isochrones.   

It is worth noting that the five stars in Figure 5 with the smallest photometric
errors (V$_o$ between 22.0 and 22.6 with blue colors) are all located redward of the 
isochrones being displayed.   This could be due to localized enhancement in the 
internal reddening in Leo~P at their locations, or to the fact that these five stars are
older than 10--20 Myr.    This latter possibility would mean that they are 
likely to be post-MS stars: blue giants rather than blue MS stars.    We will return
to this issue in \S 4.1.2.

We over-plot the CMDs with RGB isochrones  in Figure 6.  The model
data plotted here are again from the Padova group (Girardi et al.\ 2002).
In this case, we plot isochrones for models that have the same abundance 
as in Figure 5 (2\% solar), but with the age set to 12.6 Gyr.   Hence, these isochrones
should show the locations of old stellar populations in the CMDs.   We plot
both the B$-$V and V$-$R CMDs in Figure 6, since the latter shows the
putative RGB in Leo~P with much better definition.   We exhibit the model RGBs
for the same four distances used in Figure 5: 0.5, 1.0, 1.5, and 2.0 Mpc.   
The model isochrones with distances between 1.5 and 2.0 Mpc appear to 
provide the best match for the observations.  We discuss our distance 
determination methods in the following section.

\begin{figure}
\centering
\includegraphics[width=3.4in]{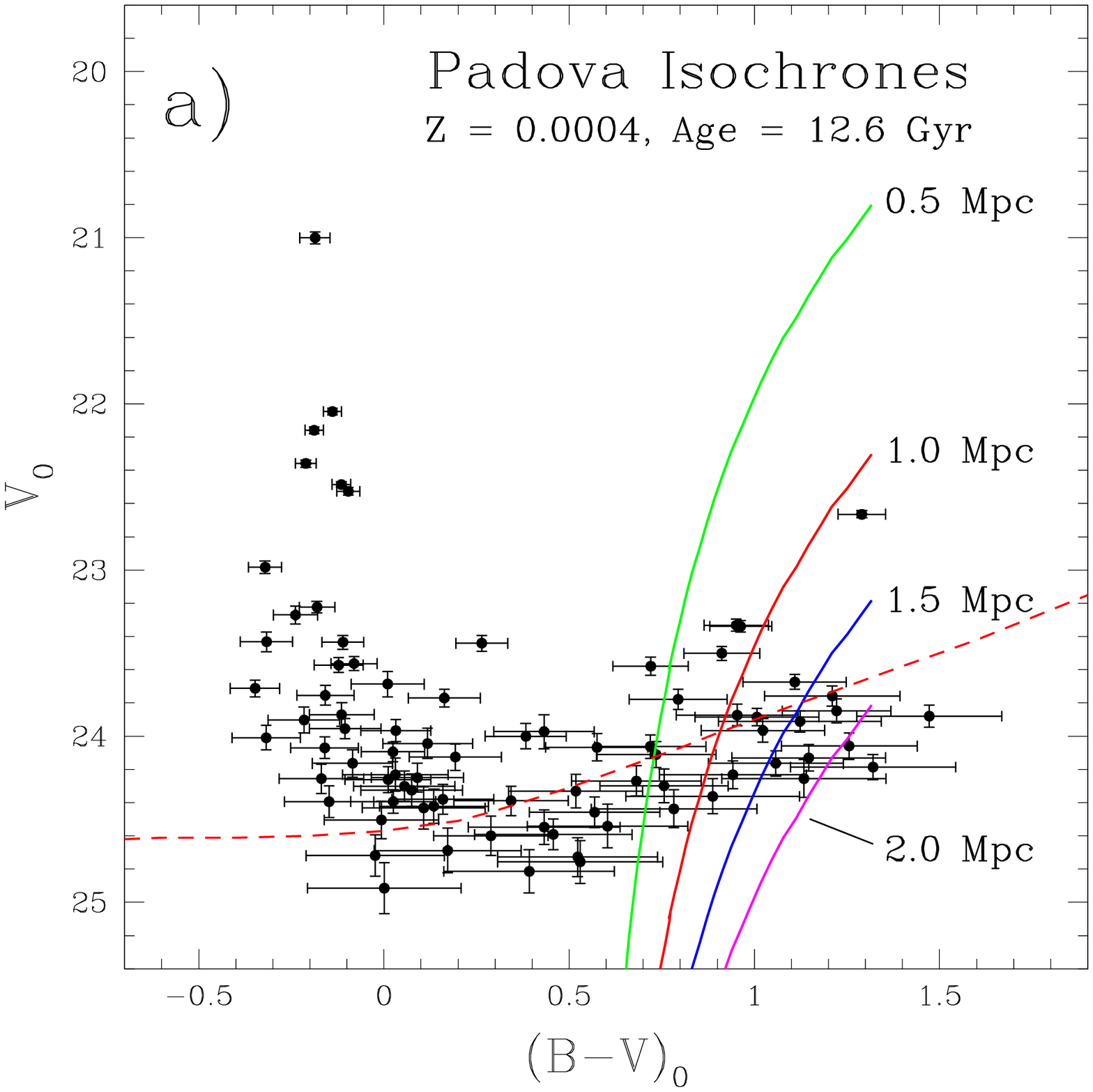}\hskip 0.1in \includegraphics[width=3.4in]{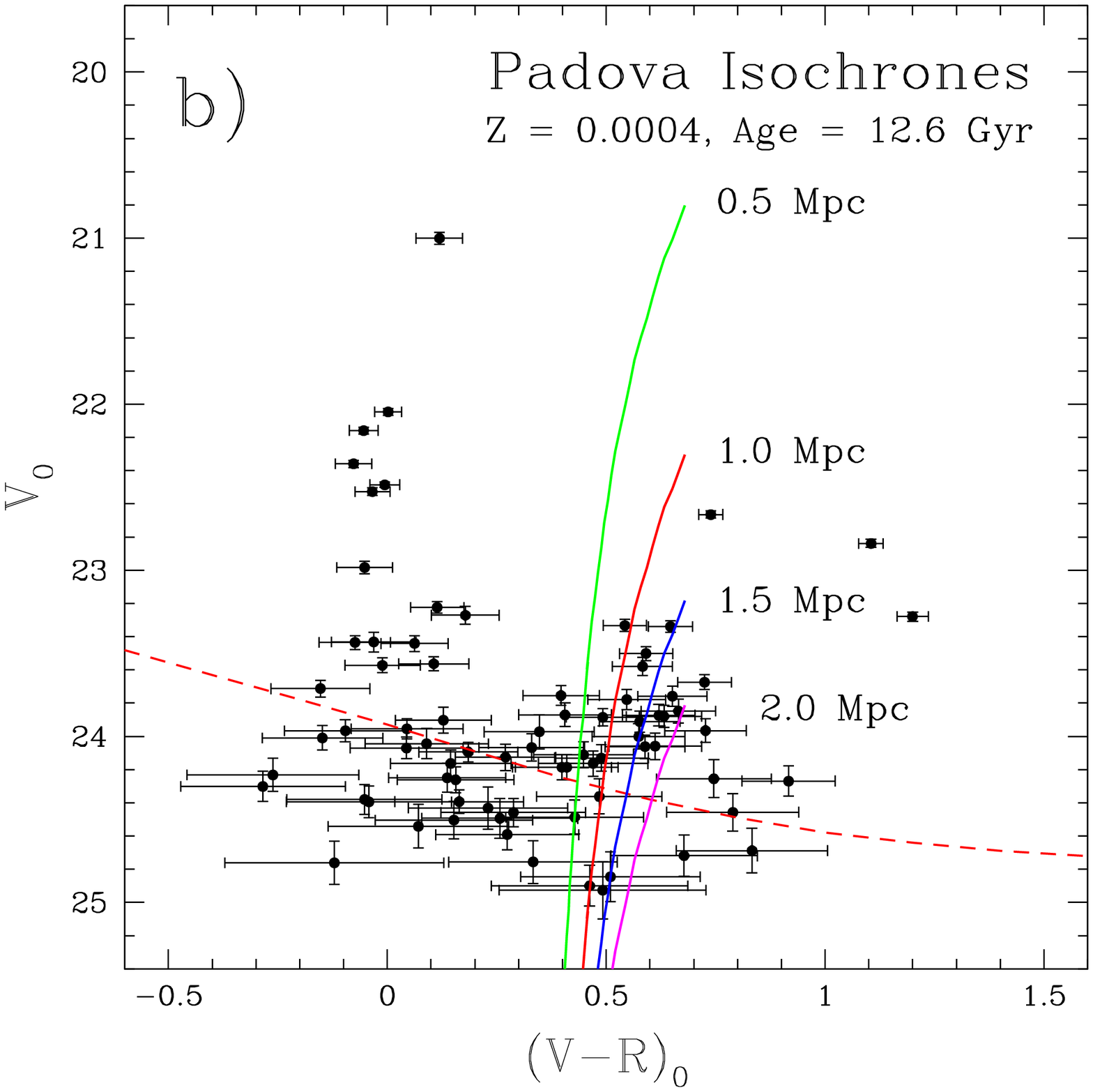}
\caption{Color-magnitude diagrams of the resolved stars in Leo~P with 
red giant branch isochrones superposed.   The data are the same as in
Figure 3, where the V$-$R CMD in panel (b) does a better job of revealing the
RGB.   The isochrones are from the Padova group (Girardi et al.\ 2002), and
show the locations of the RGB for a 12.6 Gyr old population.    We use model
isochrones with the same metallicity as those shown in Figure 5 (2\% solar).
RGBs for four different distances are plotted and labeled, using the same color
scheme employed in Figure 5.   The isochrones appear to best match the observed
RGB in the V$-$R CMD for distances between 1.5 and 2.0 Mpc.}
\end{figure}

In addition to the photometry of the individual stars, we carried out
aperture photometry of the entire galaxy.  We first masked bright
foreground stars that might contaminate our measurements, then
measured the flux in a 100\arcsec\ diameter circular aperture that
encompassed the galaxy.  The total apparent magnitude of the system 
is V$_o$ = 16.89 $\pm$ 0.01 and the broadband colors are (B$-$V)$_o$ =
0.36 $\pm$ 0.02 and (V$-$R)$_o$ = 0.49 $\pm$ 0.01.  Based on Figure 4,
we expect the foreground contamination of our total magnitudes and
colors to be quite small.   For example, if the light of all the stars present in the
comparison fields shown in Figures 4c and 4d was removed from the total aperture
magnitudes listed above, the corrected V$_o$ magnitude would change by only
0.014 and 0.013 mag, respectively.

\section{Discussion}

\subsection{Distance Estimate}

Distances to most galaxies within $\sim$10 Mpc of the Milky Way can be
determined with good accuracy using the tip of the red giant branch
(TRGB) method (e.g., Lee, Freedman \& Madore 1993; Sakai, Madore \&
Freedman 1996).  For galaxies with distances of 1 Mpc or less (i.e.,
LG members), the TRGB is located at V$_o$ $\sim$ 22.5 (Salaris \& Girardi
2005) or brighter, and accurate distances can be derived using
ground-based telescopes.  For galaxies at distances of 2 Mpc or
greater, the TRGB will be located at V$_o$ magnitudes fainter than
$\sim$24.0, and space-based observations are typically required.

In this section we describe our efforts to use the current observational data 
to constrain the distance to Leo~P.   While our data do not allow us to
determine a definitive distance, we are able to show that the galaxy is most 
likely located between 1.5 and 2.0 Mpc from the Milky Way.

\subsubsection{TRGB Method}

As mentioned in the previous section, the CMD we obtained for Leo~P is
rather unusual, especially when compared with the CMDs of other nearby
dwarf galaxies.  In particular, the upper MS is much better defined
than the RGB.  The latter is the dominant feature in most LG dwarf
galaxy CMDs (e.g., Tolstoy et al.\ 1998; Bellazzini et al.\ 2004;
Irwin et al.\ 2007; Sand et al.\ 2009).   The TRGB method relies on the 
detection of a fully populated giant branch to allow \ for the accurate 
assessment of the brightness of the RGB tip.  Our WIYN observations of Leo~P 
appear  to detect the RGB in this system, but the photometric depth is not
sufficient to provide a robust measurement of the TRGB.
Hence, any TGRB distance we derive can only be taken as approximate.  

For metal-poor stellar populations, the TRGB is located at (B$-$V)$_o$ $\sim$
1.3 (Ferraro et al.\ 1999; Stetson et al.\ 2005, 2011).  The location
of the star in Figure 6a with V$_o$ = 22.66 and (B$-$V)$_o$ = 1.29 (Star 7 in
Table 1) is consistent with it being at or near the TRGB.  Its
location in Figure 6b ((V$-$R)$_o$ = 0.74) further supports this
interpretation.  If we assume that Star 7 is in fact at the TRGB for
Leo~P, and using M$_V(TRGB)$ = $-$2.5 for the most metal-poor RGB
stars (Salaris \& Girardi 2005), the inferred distance would be D = 1.1 Mpc. 
We note, however, that associating the TRGB with this particular star then
requires that the RGB in this galaxy is under-populated in a dramatic way,
since the next star along the RGB lies nearly 0.7 mag below the tip.   At a
distance of 1.1 Mpc, Leo~P would have an absolute magnitude of $-$8.3 and
an approximate stellar mass of 1.5 $\times$ 10$^5$ M$_\odot$.   It would
seem highly unlikely that a stellar population with this mass could possess
such an under-populated RGB.   No stellar systems with an old population, 
either globular clusters or dwarf galaxies, are known that are this
massive and yet do not have a fully populated RGB (e.g., Okamoto et al. 2012).

A more likely possibility is that Star 7 is a foreground star or a red supergiant
rather than a red giant in Leo~P.  While its location in our CMDs is
precisely where one would expect a metal-poor RGB star, it is located
very far above the next brightest RGB star (Star 11, with V$_o$ = 23.33).
We can test whether Star 7 is likely to be a foreground contaminant by using the
data present in Figure 4b.  By counting the number of objects that appear
in specific magnitude and color ranges of the CMD,  then normalizing
that number by the ratio of the area covered by the full comparison
field (with Leo~P removed) to the area covered by Leo~P only, we arrive
at an estimate of the number of contaminating objects that might
appear in specific magnitude and color regions of the Leo~P CMD.
Applying this analysis to the area occupied by Star 7 leads to an estimate
that one would expect to find a foreground star with a color and brightness
comparable to Star 7 in a random field the size of Leo~P about 14\% of the time.   
This is large enough that one would not rule out the possibility of Star 7 being 
a foreground star.   Alternatively, and perhaps more likely, Star 7 could be 
located in Leo~P but be an AGB star or possibly a red supergiant star from an 
intermediate  age stellar population.

We note in passing that
there are two other very red stars located in the Leo~P field with (B$-$V)$_o$ $>$
2.1 that we believe are foreground stars.  These two stars appear in
Figure 6b as having (V$-$R)$_o$ $>$ 1.1 (they are off the plot to the right
in Figure 6a).  
Given the presence of these two likely foreground
objects in the Leo~P field, it seems at least conceivable that Star 7
may also fall in this category.    

If we assume that Star 7 is not located at the TRGB, we focus next
on the group of stars located below V$_o$ = 23.3 and with (V$-$R)$_o$
colors between 0.5 and 0.7 in Figure 6b.    The brightest of these is
Star 11 in Table 1, with V$_o$ = 23.33, (V$-$R)$_o$ = 0.54, and 
(B$-$V)$_o$ = 0.95.    Associating this star with the tip of the RGB
leads to an implied distance of 1.5 Mpc.    We note that the
B$-$V colors of Star 11 and the several stars just below it in Figure 6a 
appear to be too blue for them to located on the RGB if the distance
were closer to 1.5 Mpc.   However, in Figure 6b, where the RGB shows
better definition due to the smaller photometric errors of the red stars in 
V$-$R , these same stars appear to be consistent with an RGB with a
distance anywhere between 1.5 and 2.0 Mpc.

One should ask at this point whether or not the putative upper RGB is
heavily contaminated by foreground stars in our data.   Following the
same analysis method described above, we determined the average 
surface density of stars located in the color and brightness regime occupied 
by the top of the RGB in Leo~P.    For the magnitude range V$_o$ = 23.25 to 
24.0 and the (B$-$V)$_o$ color range 0.75 to 1.25  (the TRGB portion of the 
CMD), there are 0.71 stars per Leo~P field in our WIYN images.  The Leo~P 
CMD has 11 objects in this same region, so the estimated contamination rate 
for the upper RGB region of the Leo~P CMD is approximately 6\%.   Hence
it would appear that one can trust that the current data are not too badly
contaminated by foreground objects in the TRGB region of the CMD.

In order to derive a valid TRGB distance, we would require data that are
roughly one magnitude deeper than our current photometry.   The definition
of the RGB present in our current CMD is simply not sufficient to be able to run 
any sophisticated edge-finding software typically used for ascertaining an
accurate TRGB distance.  Our estimates above are based on single stars
which themselves might be red supergiants or AGB stars.   Therefore, it 
appears that the most robust distance estimate that we can derive for Leo~P
based on the RGB in our CMDs is 1.5 - 2.0 Mpc.  The lower end of this range 
is established by the location of the brightest star in Figure 6
that can reasonably be associated with the RGB.  The upper end is somewhat
softer, but is based on the location of the isochrones in both Figures 6a and 6b
in color space.    The vast majority of the putative RGB stars in the two CMDs
are located blueward of the 2.0 Mpc isochrone.   Since the isochrones we
are using in these figures are matched to the measured metal abundance
of Leo~P  (Skillman \etal 2013), and since differential reddening in the galaxy
will only act to push stars to redder colors in the CMDs, we infer that the
2.0 Mpc isochrone indicates a reasonable upper limit for the distance to Leo~P.

Recent ground-based imaging observations (K. McQuinn 2013, private
communication) have been obtained that reach substantially deeper than
our WIYN images.  The RGB is definitively detected in these images at
the location indicated by our data, and a very preliminary analysis of
these new data produce a distance determination that is consistent
with our range quoted above.  A thorough analysis and presentation of
these new data will be forthcoming.

\subsubsection{Distance Based on the Bright Blue Stars}

Due to the lack of a definitive distance determination from the TRGB method, we
explored an alternative approach to estimating the distance.  As we show below,
it is possible to take advantage of the single \hii\ region present in Leo~P to
constrain the distance to the galaxy.    In the case of this system, nature has provided us with
a rather unusual circumstance in which (1) there is one and only one \hii\ region in
the galaxy, and (2) the resolved stellar population of luminous blue stars is well 
delineated by our current observations.

Stars with effective temperatures cooler than $\sim$20,000 K, corresponding to 
a spectral type of B1 or B2, are not capable of producing a significant volume 
of ionized gas (Osterbrock 1989).  The brightest star in Leo~P  (Star 1) is 
surrounded by an \hii\ region, while the second 
brightest (second star in Table 1; hereafter Star 2) is not.  Our narrow-band 
images are sensitive enough to rule out any significant emission around
any of the other bright blue stars in Leo~P.   For example, if Leo~P were located
at a distance of 1.5 Mpc (e.g., consistent with the TRGB distance found in \S 4.1.1),
we would be able to detect an \hii\ region with
an H$\alpha$ luminosity of only 1.3 $\times$ 10$^{34}$ erg/s, which is 0.1\% the 
H$\alpha$ luminosity of the Orion Nebula (O'Dell, Hodge \& Kennicutt 1999).   
Hence, we can rule out with a high degree of confidence that there is any 
significant H$\alpha$ emission associated with any of the blue stars in
Leo~P other than Star 1. 

We can use the information in the paragraph above to constrain the distance
to Leo~P.  Since it has an \hii\ region around it, Star 1 (V$_o$  = 21.00) must 
have a spectral type of B2 or hotter, while Star 2 (V$_o$ = 22.05) must be B3 
or cooler.  The apparent magnitude difference between these two stars 
($\Delta$V$_o$ = 1.05 mag) further constrains their possible spectral types and 
luminosities.  For example, assuming that Star 2 is a B3V star (M$_V$ = $-$2.00), 
the observed value of $\Delta$V$_o$ would imply that Star 1 is a B0V or O9.5V  
star (M$_V$ = $-$2.90 to $-$3.05; all absolute magnitudes for O and B stars are 
taken from Wegner 2000, using the ``smoothed" values).  The implied distance 
to Leo~P in this scenario is then D = 650 kpc.   The coolest spectral class that Star 1 
could have and still possess an \hii\ region is B2V (M$_V$ = $-$2.30).  The observed 
value of $\Delta$V would then lead to Star 2 having an absolute magnitude of 
M$_V$ = $-$1.25 and a spectral type of B5V or B6V.  In this case the implied 
distance is D = 460 kpc).

The distances derived above, 460 to 650 kpc, are substantially lower than
what we found based on the TRGB method in the previous section.  Examination
of Figures 6a and 6b appear to rule out distances as small as $\sim$500 kpc with 
high confidence.    Why is our distance estimate based on the luminous blue stars
so incongruous with the TRGB distance?

Three key assumptions were made in the preceding calculations: (1) the bright blue stars
are single stars, (2) they are all on the main sequence, and (3) there is gas in the 
immediate vicinity of these stars that could be ionized if the surface temperatures 
of the stars were hot enough.   In reality, all three of these assumptions could be
incorrect.  For example, the stars may not be single stars.  It is not uncommon for massive 
stars to be formed  in binary  or multiple-star systems (e.g., all four of 
the Trapezium stars in the Orion Nebula Cluster are multiple-star systems).   If we assume
that Stars 1 and 2 are both in equal-mass binaries, the inferred distance range
would increase to 650 -- 910 kpc.

The second assumption -- that all the bright blue stars are on the main sequence --
might also be erroneous.   Recall that while Star 1 is located close to the 10 Myr isochrones
in Figure 5, the five blue stars below it (with V$_o$ between 22.0 and 22.6),
including Star 2, are all located to the right of the MS isochrones.   These five
stars may well be post-MS stars.   If this is true, then they may be
significantly older than Star 1.   Under the assumption that Star 2 is a B3 giant (B3III)
rather than a MS star, the inferred distance range for Leo~P increases to 590 -- 850 kpc
if Stars 1 and 2 are both single, and to 830 kpc - 1.20 Mpc  if they are equal-mass
binaries.   Alternatively, if Star 2 were a B3 bright giant  (B3II), the corresponding distance
range is 780 kpc -- 1.15 Mpc for single stars, and 1.10 -- 1.62 Mpc for equal-mass
binaries.   Hence, by simply invoking the assumption that the bright blue stars are
evolved stars rather than MS stars, one can derive distances that are consistent with
the TRGB distance.    Unfortunately, the precise nature of these bright blue stars is
unknown, meaning that this method is subject to unreasonably large uncertainties.

Finally, we point out that this distance estimation method relies on the assumption that
there is a significant amount of gas in close proximity to the bright blue stars that is capable
of being ionized.   While this is clearly the case for Star 1, it may not be a valid assumption 
for the other blue stars, like Star 2.   In particular, if the five stars located between V$_o$ = 
22.0 - 22.6 with blue colors are all post-MS stars, as suggested in \S 3 and inferred above,
then it would seem likely that they have pushed away the gas associated with their formation
episode via stellar winds.   It is even possible that they are located in small associations 
that once included higher-mass stars which ended their lives as supernovae and therefore 
cleared out the ISM in the immediate vicinity of the groups.

It seems clear that we do not know enough about these individual stars to be able to
generate a reliable distance from them using this method.   The fact that the luminosity classes 
of these stars are unknown,  combined with the possibility of binarity or multiplicity, yields
distances that range over a factor of nearly four (460 kpc to 1.62 Mpc).   It is nevertheless
reassuring that the range covered by these possible distances overlaps the distance
range inferred from the TRGB method.

\subsubsection{Summary of Distance Estimates}

We summarize the current status of the distance estimate for Leo~P as
being uncertain.   We adopt a distance of 1.5 to 2.0 Mpc, based on the 
apparent location of the tip of the RGB in our CMDs.   Our attempt to use
the bright blue stars that have robust photometry to constrain the distance
leads to a distance range that is unacceptably large, although it is consistent
with our TRGB estimate under the assumption that at least some of the blue stars present
in the galaxy are post-MS objects.   We conclude that an unequivocal distance 
determination is not possible with the current data, and that deeper high-resolution 
imaging data are required to arrive at a more precise value of the distance.  Such
data have been obtained recently (K. McQuinn, private communication), and appear
to corroborate our distance estimate.

It is appropriate to note that Giovanelli \etal (2013) used a completely 
independent method to estimate the distance to Leo~P.   Using the baryonic
Tully-Fisher relation (e.g., McGaugh 2012) and the observed rotation
velocity of the galaxy derived from their HI data, Giovanelli \etal derive a
distance  to Leo~P of 1.3 (+0.9, -0.5) Mpc.   This value is consistent with
our distance estimates derived using the TRGB method.

\subsection{Galaxy Properties}

Although our distance estimate for Leo~P is uncertain, it is
nevertheless a useful exercise to derive and consider the physical
properties of this newly discovered galaxy.  For this exercise we
assume the range of distances based on our TRGB estimates: 1.5 - 2.0
Mpc.  We present a number of observed and derived quantities in Table
2 computed for three different distances: 1.5, 1.75, and 2.0 Mpc.  The
observed quantities, obtained from the optical observations presented
in the current paper, include the right ascension and declination of
the \hii\ region in Leo~P, the integrated V$_o$ magnitude plus
(B$-$V)$_o$ and (V$-$R)$_o$ colors and their errors, and the total
H$\alpha$ flux of the \hii\ region.  We use the assumed distances to
compute the absolute magnitude, optical diameter, and stellar and
\hi\ mass estimates for the galaxy.  Stellar masses are derived using
the methodology of Bell \& de Jong (2001), and the \hi\ mass estimate
uses the \hi\ flux from Giovanelli et al.\ (2013).  We also derive the
H$\alpha$ luminosity of the single \hii\ region present in Leo~P after
applying the Schlegel et al. (1998) correction for Galactic absorption
to our observed H$\alpha$ flux.

\begin{deluxetable*}{lccc} 
\tabletypesize{\scriptsize}
\tablecaption{Properties of Leo~P \label{tab:parms}}
\tablehead{
\colhead{Parameter}&& \colhead{Value}    
}
\startdata
RA -- \hii\ region (J2000)	 && 10:21:45.1 	\\
DEC -- \hii\ region (J2000)\ \ \ && 18:05:17.2   	\\
\\
V$_o$ magnitude	 	&& 16.89 $\pm$ 0.01	\\
(B$-$V)$_o$ color		&& 0.36 $\pm$ 0.02  \\
(V$-$R)$_o$ color		&& 0.49 $\pm$ 0.01	\\
F$_{H\alpha}$ [erg/s/cm$^2$]	&& (1.71 $\pm$ 0.03) $\times$ 10$^{-14}$  	\\
\\
\hline
\\
&& \hskip -1.99in Distance-Dependent Properties: \\
\\
Distance D (assumed) [Mpc]  & 1.5 & 1.75 & 2.0 \\
M$_V$		& $-$9.0 & $-$9.3 & $-$9.6      \\
Diameter	[pc]	& 650 & 760 & 870  \\
L$_{H\alpha}$ [erg/s]	& 4.8 $\times$ 10$^{36}$  & 6.6 $\times$ 10$^{36}$ & 8.6 $\times$ 10$^{36}$\\
L$_V$ [L$_\odot$]		&  331,000 & 450,000 & 588,000	\\
Stellar Mass [M$_\odot$]	& 269,000 & 366,000 & 479,000	\\
\hi\ Mass [M$_\odot$]	& 689,000 & 937,000 & 1,224,000	\\
\enddata
\end{deluxetable*}

The physical properties of Leo~P listed in Table 2 make it clear that
this is an interesting system.  If we adopt for this discussion a
distance of 1.75 $\pm$ 0.25 Mpc, then the picture that emerges is one
of a very low-luminosity dwarf galaxy (M$_V$ = $-$9.3 $\pm$ 0.3;
diameter = 760 $\pm$ 90 pc) that is also extremely gas rich.  It has
an \hi\ mass of 9.4 $\times$ 10$^5$ M$_\odot$ and an \hi-to-stellar
mass ratio M$_{HI}$/M$_*$ $\sim$ 2.6 (the latter number being
independent of the distance).  It is significantly less luminous than
other gas-rich dwarfs like Leo A and Phoenix (M$_V$ $=$ $-$12.1 and
$-$9.9, respectively; McConnachie 2012), and hence represents the
least luminous/massive system in the local universe known to have
current star formation.  The star formation rate (SFR) implied by the
H$\alpha$ luminosity is log(SFR) $=$ $-$4.27, assuming the Kennicutt
(1998) conversion factor.  This value, while quite low in an absolute
sense, places Leo~P slightly {\it above} the observed trend line in
the SFR$-$L$_B$ relation derived from H$\alpha$ fluxes of galaxies in
the Local Volume (e.g., Lee et al. 2009; Karachentsev \& Kaisin 2010).
By comparison, only one other galaxy with M$_V$ fainter than
approximately $-$10.6 in the two samples cited above shows evidence
for current star formation (as indicated by detected H$\alpha$
emission).

While the current ground-based photometric data do not allow for a
detailed analysis, we can still infer much useful information about
the stellar populations present in Leo~P.  The optical appearance
(Figure 1) is dominated by the young, bright blue stars that are
clustered in the southern portion of the galaxy.  The implied age of
Star 1, the central star in the single \hii\ region, is no more than
10--15 Myr assuming that it is a MS star.  It seems possible that at
least some of the bright blue stars seen in Leo~P are older than Star
1 and have evolved off of the MS.  These stars could be more than 30
Myr old.  This age makes it possible that they were created in an
earlier phase of the same event that spawned Star 1, i.e., there may
well be linkage between the star-formation events that created all of
the luminous blue stars currently seen in Leo~P.  Whether there is a
second intermediate-age population (e.g., with an age of a
few-to-several hundred Myr) is unclear from the current data.  There
was certainly star formation in the distant past in Leo~P, since there
is clearly a population of red giants present.

One can speculate about what Leo~P will look like in the optical
in $\sim$500 Myr, when all of the bright blue stars will have ended
their stellar lives.  At that point, and assuming no additional star 
formation takes place, the southern high-surface-brightness portion 
of the galaxy will more nearly resemble the current northern 
low-surface-brightness region.  It is not clear whether Leo~P would be 
recognizable (or detectable) as a galaxy in the optical after 500 Myr, 
especially if star formation were to cease.    Conversely, what did
Leo~P look like several tens of millions of years in the past?   The
degree of brightening associated with the recent episode of star formation
suggests that there could be other objects similar to Leo~P  in the
local universe lying just below the level of easy detectability at
optical wavelengths.  Future deep searches of  apparently starless \hi\ clouds of
the type detected by ALFALFA (Giovanelli et al.\ 2010, 2013) might well
uncover additional examples of low-stellar-mass  dwarf galaxies.

The single \hii\ region in Leo~P has an H$\alpha$ luminosity of
roughly 50\% that of the Orion Nebula (O'Dell, Hodge \& Kennicutt
1999) if the 1.75 Mpc distance is assumed.  This luminosity is 
consistent with the ionizing flux being contributed by a single O7 or O8 type
star.  Our group has obtained spectroscopy of this \hii\ region that
reveals a strong emission-line spectrum indicative of a very
metal-poor ISM.  A full nebular analysis provides an estimate of the
oxygen abundance of log(O/H)+12 $\sim$ 7.1--7.2 ($\sim$2\% of the solar
value), making Leo~P one of the lowest metallicity galaxies known.
The details of the spectral analysis of Leo~P are presented by
Skillman \etal\ (2013).

Despite our incomplete understanding of Leo~P, it is still possible to
speculate about the evolutionary status of this system.  It would
appear likely that this galaxy has not passed too close to the
Milky Way or Andromeda (yet), since otherwise it is difficult to understand
how it could have retained its gas despite its extremely low mass.
Whether Leo~P has maintained this amount of gas mass since its initial
collapse or has accreted most of it more recently is a key question.
Its apparent location outside of the LG places it in a relatively benign
environment, far beyond the virial radius of either of the giant LG
galaxies.  Given this location, what process can be invoked to explain
its recent episode of star formation?  It would appear to be
relatively isolated, unless its distance is much lower than we have
estimated in the current paper.  We note in passing that Leo~P is located 
close on the sky ($\sim$7$^o$ away) to Leo I.  Furthermore, Leo I has an
observed velocity very similar to Leo~P (285 km/s and 264 km/s,
respectively).   However, Leo I has a secure distance determination of $\sim$250
kpc (Bellazzini et al. 2004), which rules out any physical connection between the
two.  According to McConnachie (2012), Leo I is possibly not gravitationally bound 
to the Milky Way, but its location and velocity are consistent with it
being bound to the LG.  Despite its low velocity of 137 \kms\ in the LG
reference frame (Giovanelli et al. 2013), the inferred distance to Leo~P 
makes it unlikely that it is bound to the LG.   It would appear that its relative isolation
has allowed Leo~P to survive and retain its gas since its initial formation.    
Its relatively pristine nature seems likely to be an outcome of this isolation.

\section{Summary}

We have presented optical observations of the dwarf galaxy Leo~P,
which was discovered by the ALFALFA \hi\ survey (Giovanelli et
al.\ 2013).  Star-forming activity is ongoing in Leo~P, as revealed by
the detection in our data of a number of luminous blue stars and an \hii\ region
apparently ionized by a single star.  Those observations and evidence
of a RGB constrain the distance to the galaxy to be between 1.5 and
2.0 Mpc.  The definition of the upper end of the RGB in our data is poor, 
making a precise distance determination impossible with the current 
data set.  Additional observations will be required to more accurately 
characterize Leo~P's RGB and to obtain a definitive distance.  The limits 
provided by our data are sufficient to indicate that Leo~P is probably
located outside the LG in the Local Volume. Those limits and the
measurement of an ultra-low metallicity (Skillman \etal\ 2013)
indicate that Leo~P is an extreme object which may well have the
lowest mass and luminosity of any star-forming galaxy known.  The
ALFALFA survey is finding other HI sources that resemble Leo~P (Adams
\etal 2013), providing an exploratory entry to the
regime of the lowest mass, most pristine galaxies.

\acknowledgments 

Support for KLR and MDY was provided by an NSF Faculty Early Career
Development (CAREER) award (AST-0847109; PI: Rhode).  The ALFALFA team
at Cornell is supported by NSF grants AST-0607007 and AST-1107390 to
RG and MPH and by grants from the Brinson Foundation.  EAKA is
supported by an NSF predoctoral fellowship. JMC is supported by NSF
grant AST-1211683.  We thank the staff of the WIYN Observatory and
Kitt Peak National Observatory for their assistance with obtaining the
optical data for this study.  We acknowledge the insightful comments
from an anonymous referee that greatly improved this paper.  This
research has made use of the NASA/IPAC Extragalactic Database (NED)
which is operated by the Jet Populsion Laboratory, California
Institute of Technology, under contract with the National Aeronautics
and Space Administration.




\end{document}

%% file: table_1.dat
1 & 10:21:45.1 & $+$18:05:17.3 & 21.00 & 0.04 & $-$0.19 & 0.04 & 0.12 & 0.05 \\
2 & 10:21:45.0 & $+$18:05:25.6 & 22.05 & 0.02 & $-$0.14 & 0.03 & 0.00 & 0.03 \\
3 & 10:21:44.9 & $+$18:05:23.1 & 22.16 & 0.02 & $-$0.19 & 0.03 & $-$0.05 & 0.03 \\
4 & 10:21:44.8 & $+$18:05:23.0 & 22.36 & 0.02 & $-$0.21 & 0.03 & $-$0.08 & 0.04 \\
5 & 10:21:43.9 & $+$18:05:57.3 & 22.48 & 0.02 & $-$0.12 & 0.03 & $-$0.01 & 0.03 \\
6 & 10:21:45.0 & $+$18:05:26.3 & 22.53 & 0.02 & $-$0.10 & 0.03 & $-$0.03 & 0.04 \\
7 & 10:21:44.8 & $+$18:05:31.6 & 22.66 & 0.02 & 1.29 & 0.07 & 0.74 & 0.03 \\
8 & 10:21:45.1 & $+$18:05:14.5 & 22.98 & 0.04 & $-$0.32 & 0.04 & $-$0.05 & 0.06 \\
9 & 10:21:44.8 & $+$18:05:27.8 & 23.22 & 0.04 & $-$0.18 & 0.05 & 0.11 & 0.06 \\
10 & 10:21:45.0 & $+$18:05:14.0 & 23.27 & 0.05 & $-$0.24 & 0.06 & 0.18 & 0.08 \\
11 & 10:21:45.5 & $+$18:05:13.7 & 23.33 & 0.04 & 0.95 & 0.09 & 0.54 & 0.05 \\
12 & 10:21:44.7 & $+$18:05:48.8 & 23.34 & 0.04 & 0.96 & 0.08 & 0.65 & 0.05 \\
13 & 10:21:44.7 & $+$18:05:24.3 & 23.43 & 0.06 & $-$0.32 & 0.07 & $-$0.03 & 0.10 \\
14 & 10:21:45.2 & $+$18:05:21.0 & 23.43 & 0.04 & $-$0.11 & 0.06 & $-$0.07 & 0.08 \\
15 & 10:21:44.0 & $+$18:05:24.5 & 23.44 & 0.05 & 0.26 & 0.07 & 0.06 & 0.08 \\
16 & 10:21:44.7 & $+$18:05:34.9 & 23.50 & 0.04 & 0.91 & 0.10 & 0.59 & 0.06 \\
17 & 10:21:44.6 & $+$18:05:20.6 & 23.56 & 0.04 & $-$0.08 & 0.06 & 0.11 & 0.08 \\
18 & 10:21:45.0 & $+$18:05:06.9 & 23.57 & 0.04 & $-$0.12 & 0.07 & $-$0.01 & 0.09 \\
19 & 10:21:43.4 & $+$18:05:38.8 & 23.58 & 0.05 & 0.72 & 0.10 & 0.58 & 0.07 \\
20 & 10:21:44.9 & $+$18:05:38.7 & 23.67 & 0.04 & 1.11 & 0.14 & 0.72 & 0.06 \\
21 & 10:21:44.5 & $+$18:05:32.6 & 23.69 & 0.08 & 0.01 & 0.10 & ... & ...\\
22 & 10:21:44.7 & $+$18:05:21.8 & 23.71 & 0.05 & $-$0.35 & 0.07 & $-$0.15 & 0.11 \\
23 & 10:21:45.2 & $+$18:05:09.8 & 23.75 & 0.06 & $-$0.16 & 0.08 & 0.40 & 0.09 \\
24 & 10:21:43.9 & $+$18:05:52.3 & 23.76 & 0.06 & 1.21 & 0.18 & 0.65 & 0.08 \\
25 & 10:21:45.2 & $+$18:05:07.6 & 23.77 & 0.05 & 0.16 & 0.10 & ... & ...\\
26 & 10:21:43.6 & $+$18:05:41.7 & 23.78 & 0.06 & 0.79 & 0.13 & 0.55 & 0.09 \\
27 & 10:21:44.2 & $+$18:05:58.6 & 23.85 & 0.07 & 1.22 & 0.15 & 0.67 & 0.08 \\
28 & 10:21:45.2 & $+$18:05:13.3 & 23.87 & 0.07 & $-$0.11 & 0.09 & 0.41 & 0.11 \\
29 & 10:21:44.2 & $+$18:05:53.2 & 23.87 & 0.07 & 0.95 & 0.16 & 0.62 & 0.08 \\
30 & 10:21:45.1 & $+$18:05:30.1 & 23.88 & 0.07 & 1.47 & 0.20 & 0.63 & 0.09 \\
31 & 10:21:45.7 & $+$18:04:57.2 & 23.89 & 0.05 & 1.01 & 0.17 & 0.49 & 0.08 \\
32 & 10:21:45.0 & $+$18:05:10.5 & 23.90 & 0.08 & $-$0.22 & 0.10 & 0.13 & 0.11 \\
33 & 10:21:45.3 & $+$18:05:22.4 & 23.91 & 0.06 & 1.12 & 0.22 & 0.58 & 0.09 \\
34 & 10:21:44.7 & $+$18:05:16.9 & 23.95 & 0.06 & $-$0.10 & 0.10 & 0.04 & 0.13 \\
35 & 10:21:43.1 & $+$18:06:02.6 & 23.96 & 0.07 & 1.02 & 0.17 & 0.73 & 0.09 \\
36 & 10:21:44.5 & $+$18:05:26.4 & 23.97 & 0.07 & 0.03 & 0.09 & $-$0.10 & 0.14 \\
37 & 10:21:43.6 & $+$18:05:55.2 & 23.97 & 0.10 & 0.43 & 0.14 & 0.35 & 0.13 \\
38 & 10:21:43.0 & $+$18:05:59.8 & 24.00 & 0.08 & 0.38 & 0.11 & 0.57 & 0.11 \\
39 & 10:21:45.0 & $+$18:05:16.1 & 24.01 & 0.07 & $-$0.32 & 0.09 & $-$0.15 & 0.14 \\
40 & 10:21:43.8 & $+$18:05:26.1 & 24.04 & 0.09 & 0.12 & 0.12 & 0.09 & 0.14 \\
41 & 10:21:44.4 & $+$18:05:34.1 & 24.06 & 0.08 & 1.26 & 0.18 & 0.61 & 0.11 \\
42 & 10:21:45.3 & $+$18:05:06.4 & 24.06 & 0.07 & 0.72 & 0.15 & 0.59 & 0.09 \\
43 & 10:21:44.0 & $+$18:05:19.0 & 24.07 & 0.08 & 0.57 & 0.14 & 0.33 & 0.12 \\
44 & 10:21:45.3 & $+$18:05:08.5 & 24.07 & 0.07 & $-$0.16 & 0.09 & 0.04 & 0.13 \\
45 & 10:21:45.4 & $+$18:05:03.6 & 24.09 & 0.06 & 0.02 & 0.09 & 0.18 & 0.11 \\
46 & 10:21:45.2 & $+$18:05:37.6 & 24.11 & 0.08 & 0.73 & 0.16 & 0.45 & 0.11 \\
47 & 10:21:44.4 & $+$18:05:30.1 & 24.13 & 0.08 & 0.19 & 0.12 & 0.27 & 0.11 \\
48 & 10:21:43.8 & $+$18:05:52.2 & 24.13 & 0.08 & 1.15 & 0.21 & 0.49 & 0.11 \\
49 & 10:21:44.6 & $+$18:05:22.2 & 24.16 & 0.08 & $-$0.09 & 0.11 & 0.14 & 0.14 \\
50 & 10:21:46.2 & $+$18:05:05.8 & 24.16 & 0.08 & 1.06 & 0.18 & 0.47 & 0.12 \\
51 & 10:21:44.0 & $+$18:06:01.8 & 24.19 & 0.07 & 1.32 & 0.22 & 0.41 & 0.12 \\
52 & 10:21:44.4 & $+$18:05:26.0 & 24.23 & 0.10 & 0.03 & 0.14 & $-$0.26 & 0.20 \\
53 & 10:21:45.7 & $+$18:05:25.8 & 24.23 & 0.08 & 0.94 & 0.20 & ... & ...\\
54 & 10:21:45.3 & $+$18:05:30.5 & 24.25 & 0.09 & 0.09 & 0.12 & 0.14 & 0.13 \\
55 & 10:21:43.3 & $+$18:06:01.3 & 24.25 & 0.11 & 1.13 & 0.22 & 0.75 & 0.13 \\
56 & 10:21:44.7 & $+$18:05:15.5 & 24.26 & 0.09 & $-$0.17 & 0.11 & ... & ...\\
57 & 10:21:44.8 & $+$18:05:46.2 & 24.26 & 0.08 & 0.01 & 0.12 & 0.16 & 0.13 \\
58 & 10:21:45.4 & $+$18:05:31.9 & 24.27 & 0.09 & 0.68 & 0.17 & 0.92 & 0.11 \\
59 & 10:21:44.0 & $+$18:05:54.7 & 24.30 & 0.10 & 0.76 & 0.17 & ... & ...\\
60 & 10:21:45.1 & $+$18:05:11.7 & 24.30 & 0.09 & 0.06 & 0.14 & $-$0.28 & 0.19 \\
61 & 10:21:44.3 & $+$18:05:18.1 & 24.33 & 0.10 & 0.07 & 0.14 & ... & ...\\
62 & 10:21:43.0 & $+$18:05:50.2 & 24.33 & 0.10 & 0.52 & 0.18 & ... & ...\\
63 & 10:21:44.4 & $+$18:05:27.5 & 24.36 & 0.10 & 0.89 & 0.23 & 0.48 & 0.14 \\
64 & 10:21:44.6 & $+$18:05:13.7 & 24.38 & 0.09 & 0.16 & 0.14 & $-$0.05 & 0.18 \\
65 & 10:21:44.4 & $+$18:05:37.6 & 24.39 & 0.09 & 0.34 & 0.15 & ... & ...\\
66 & 10:21:45.9 & $+$18:05:21.3 & 24.39 & 0.07 & 0.03 & 0.12 & 0.16 & 0.15 \\
67 & 10:21:45.9 & $+$18:04:55.3 & 24.39 & 0.10 & $-$0.15 & 0.12 & $-$0.04 & 0.19 \\
68 & 10:21:44.6 & $+$18:05:29.3 & 24.42 & 0.10 & 0.14 & 0.15 & $-$0.85 & 0.37 \\
69 & 10:21:45.0 & $+$18:05:38.2 & 24.43 & 0.13 & 0.11 & 0.17 & 0.23 & 0.18 \\
70 & 10:21:44.8 & $+$18:05:09.4 & 24.44 & 0.11 & 0.78 & 0.22 & ... & ...\\
71 & 10:21:43.4 & $+$18:05:48.3 & 24.46 & 0.09 & 0.57 & 0.18 & ... & ...\\
72 & 10:21:44.2 & $+$18:05:16.4 & 24.50 & 0.11 & $-$0.01 & 0.15 & 0.15 & 0.18 \\
73 & 10:21:44.7 & $+$18:05:46.2 & 24.54 & 0.13 & 0.60 & 0.22 & 0.07 & 0.21 \\
74 & 10:21:44.1 & $+$18:05:34.5 & 24.55 & 0.10 & 0.43 & 0.20 & ... & ...\\
75 & 10:21:44.9 & $+$18:05:34.4 & 24.59 & 0.09 & 0.46 & 0.21 & 0.27 & 0.16 \\
76 & 10:21:44.1 & $+$18:05:40.5 & 24.60 & 0.12 & 0.29 & 0.15 & ... & ...\\
77 & 10:21:44.0 & $+$18:05:55.9 & 24.69 & 0.13 & 0.17 & 0.20 & 0.83 & 0.17 \\
78 & 10:21:45.1 & $+$18:05:39.0 & 24.72 & 0.12 & $-$0.02 & 0.19 & 0.68 & 0.17 \\
79 & 10:21:43.9 & $+$18:05:46.9 & 24.73 & 0.12 & 0.52 & 0.22 & ... & ...\\
80 & 10:21:43.7 & $+$18:05:48.6 & 24.76 & 0.13 & 0.53 & 0.22 & 0.33 & 0.19 \\
81 & 10:21:45.4 & $+$18:05:36.5 & 24.81 & 0.13 & 0.39 & 0.23 & ... & ...\\
82 & 10:21:45.0 & $+$18:05:34.4 & 24.91 & 0.15 & 0.00 & 0.21 & ... & ...\\